\documentclass[reprint,amsmath,amssymb,aps,floatfix,tightenlines,superscriptaddress]{revtex4-1}
\usepackage[latin9]{inputenc}
\usepackage{color}
\usepackage{amsmath}
\usepackage{amssymb}
\usepackage{graphicx}
\usepackage{wasysym}
\usepackage[unicode=true,
 bookmarks=false,
 breaklinks=false,pdfborder={0 0 1}]
 {hyperref}

\makeatletter

\providecommand{\tabularnewline}{\\}

\usepackage{bookmark}
\usepackage{amsfonts}
\usepackage{dcolumn}
\usepackage{bm}
\usepackage{hyperref}
\hypersetup{
	colorlinks=false,
	linkcolor=black,      
	urlcolor=black,
	citecolor=black,
}

\makeatother

\begin{document}
\title{Phenomenological model for the third-harmonic magnetic response of
superconductors: application to Sr$_{2}$RuO$_{4}$}
\author{Fei Chen}
\affiliation{School of Physics and Astronomy, University of Minnesota, Minneapolis,
55455, USA}
\author{Damjan Pelc}
\affiliation{School of Physics and Astronomy, University of Minnesota, Minneapolis,
55455, USA}
\affiliation{Department of Physics, Faculty of Science, University of Zagreb, Bijeni\v{c}ka
32, HR-10000 Zagreb, Croatia}
\author{Martin Greven}
\affiliation{School of Physics and Astronomy, University of Minnesota, Minneapolis,
55455, USA}
\author{Rafael M. Fernandes}
\affiliation{School of Physics and Astronomy, University of Minnesota, Minneapolis,
55455, USA}
\begin{abstract}
We employ the phenomenological Lawrence-Doniach model to compute the
contributions of the superconducting fluctuations to the third-harmonic
magnetic response, denoted here by $\overline{M_{3}}$, which can
be measured in a precise way using ac magnetic fields and lock-in
techniques. We show that, in an intermediate temperature regime, this
quantity behaves as the third-order nonlinear susceptibility, which
shows a power-law dependence with the reduced temperature $\epsilon=\frac{T-T_{c}}{T_{c}}$
as $\epsilon^{-5/2}$. Very close to $T_{c}$, however, $\overline{M_{3}}$
saturates due to the nonzero amplitude of the ac field. We compare
our theoretical results with experimental data for three conventional
superconductors -- lead, niobium, and vanadium -- and for the unconventional
superconductor Sr$_{2}$RuO$_{4}$ (SRO). We find good agreement between
theory and experiment for the elemental superconductors, although
the theoretical values for the critical field systematically deviate
from the experimental ones. In the case of SRO, however, the phenomenological
model completely fails to describe the data, as the third-harmonic
response remains sizable over a much wider reduced temperature range
compared to Pb, Nb, and V. We show that an inhomogeneous distribution
of $T_{c}$ across the sample can partially account for this discrepancy,
since regions with a locally higher $T_{c}$ contribute to the fluctuation
$\overline{M_{3}}$ significantly more than regions with the ``nominal''
$T_{c}$ of the clean system. However, the exponential temperature
dependence of $\overline{M_{3}}$ first reported in Ref. [D. Pelc \emph{et. al.}, Nature Comm. \textbf{10}, 2729 (2019)]
is not captured by the model with inhomogeneity. We conclude that,
while inhomogeneity is an important ingredient to understand
the superconducting fluctuations of SRO and other perovskite superconductors,
additional effects may be at play, such as non-Gaussian fluctuations
or rare-region effects. 
\end{abstract}
\date{\today}

\maketitle

\section{Introduction}

In unconventional superconductors, not only the gap function, but
also the superconducting fluctuations can be quite different from
their conventional counterparts (for reviews, see Ref. \citep{varlamov2018,larkin2005book}).
Indeed, several high-$T_{c}$ superconductors have strongly anisotropic
properties and small coherence lengths, suggestive of a wider temperature
range in which fluctuations are important. Moreover, the magnitude
of these fluctuations as well as their temperature dependence can
also display unusual behaviors \citep{pelc2019}. Signatures of superconducting
fluctuations have been widely probed in both conventional and unconventional
superconductors, in observables as diverse as specific heat \citep{suzuki1977,tsuboi1977,tallon2011},
linear and nonlinear conductivity \citep{glover1967,strongin1968,ruggiero1980,mircea2009,rullier2011,leridon2016,popcevic2018,Pelc2018},
microwave and THz response \citep{orenstein1999,orenstein2006,grbic2011,bilbro2011,Pelc2018},
susceptibility \citep{geballe1971,gollub1973,ong2010,kokanovic2013,Kasahara2016,yu2019},
and the Nernst coefficient \citep{ong2006,Forget2006,taillefer2012,taillefer2014,behnia2016}.

Experimentally, one of the main difficulties is to unambiguously identify
contributions that can be uniquely attributed to superconducting fluctuations,
since these are usually small compared to the regular normal-state
contributions \citep{geballe1971}. Theoretically, modeling contributions
of superconducting fluctuations to the magnetic susceptibility and
to the conductivity, both phenomenologically and microscopically,
dates back several decades \citep{schmidt1968onset,shmidtvv1968,maki1968,thompson1970,schmid1969,prange1970,Abrahams1970,kurkijarvi1972,Aslamazov1974}.
More recent studies on superconducting fluctuations have focused on
the role of phase fluctuations \citep{Emery1995}, on disordered 2D
superconductors \citep{glatz2011}, and on thermal and electric transport
properties above $T_{c}$ in cuprates \citep{Ullah1991,Fisher1991,loffe1993,Huse2002,Galistki09,Michaeli2009,Levchenko2020}.

Recently, a method to probe superconducting fluctuations based on
the third-harmonic magnetic response was put forward in Ref. \citep{pelc2019}.
Specifically, an ac magnetic field $H(t)=H_{0}\cos(\omega t)$ is
applied and the magnetization is measured at a frequency $3\omega$.
This observable, which we hereafter denote by $\overline{M_{3}}$,
is related to, but not identical to the standard nonlinear susceptibility
$\chi_{3}$. The key point is that the third-harmonic response $\overline{M_{3}}$
is vanishingly small in the normal state. As a result, its magnitude
and temperature dependence near the superconducting transition temperature
$T_{c}$ should be dominated by superconducting fluctuations. In Ref.
\citep{pelc2019}, it was empirically found that $\overline{M_{3}}$
displays an unusual exponential temperature dependence in perovskite-based
superconductors such as cuprates, Sr$_{2}$RuO$_{4}$ (SRO) and SrTiO$_{3}$,
as opposed to a power-law temperature dependence in standard electron-phonon
superconductors. However, the implications of these observations for
the nature of superconducting fluctuations in unconventional superconductors
remain unsettled.

In this paper, we employ a phenomenological approach based on the
Lawrence-Doniach (LD) free-energy to compute
the contributions to the experimentally-measured quantity $\overline{M_{3}}$
of Ref. \citep{pelc2019} arising from Gaussian superconducting fluctuations.
The main appeal of such an approach is that, being phenomenological,
it is potentially applicable to both conventional and unconventional
superconductors. In particular, we perform a quantitative comparison
between the theoretical results predicted by the LD formalism and
the data on several elemental superconductors (Pb, Nb, V) and on the
unconventional superconductor SRO. We find that the LD result provides
a good description of the data for elemental superconductors over
a wide range of reduced temperature values, $\epsilon\equiv\frac{T-T_{c}}{T_{c}}$,
and correctly captures the observed $5/2$ power-law behavior of $\overline{M_{3}}$
for intermediate values of $\epsilon$. The theoretically extracted
values for the zero-temperature upper critical field $H_{c2}(0)$
differ by factors of $2$ to $6$ from the experimental ones; we argue
that this difference could be an artifact of the LD model, which was developed for layered superconductors rather than cubic systems.
Overall, the results demonstrate that measurements of the third-harmonic
magnetic response are indeed a powerful probe of superconducting fluctuations.

However, in the case of Sr$_{2}$RuO$_{4}$, we find a sharp disagreement
between the LD theoretical results and the data for $\overline{M_{3}}$.
Not only is the temperature dependence qualitatively different, but
the observed magnitude of $\overline{M_{3}}$ near $T_{c}$ is strongly
underestimated by the theoretical model. Motivated by the evidence
for significant inhomogeneity in several perovskite-based superconductors
\citep{Pelc2018,pelc2019,pelc2021}, we modify our LD model for $\overline{M_{3}}$
and include a distribution of $T_{c}$ values. We find that even a
modest width of this $T_{c}$ distribution is capable of capturing
the typical values of $\overline{M_{3}}$ observed experimentally.
However, this modification is not sufficient to explain the exponential
temperature dependence reported in Ref. \citep{pelc2019}. We thus
conclude that while inhomogeneity at the mean-field level is important to
elucidate the behavior of superconducting fluctuations in Sr$_{2}$RuO$_{4}$,
it is likely not the sole reason for the observed exponential temperature
dependence. One possibility is that such behavior arises from rare-region
contributions \citep{dodaro2018,pelc2019,pelc2021} or from non-Gaussian
fluctuations, which are absent in the LD model employed here.

The paper is organized as follows: in Sec. \ref{sec:Phenomenology},
we employ the LD model to derive an expression for the third-harmonic
magnetic response $\overline{M_{3}}$, and discuss the temperature
dependence of this quantity in different regimes. Sec. \ref{sec:Comparison}
presents a quantitative comparison between the theoretical and experimental
results for three conventional superconductors (Pb, Nb, and V) and
the unconventional superconductor Sr$_{2}$RuO$_{4}$. We note that
some of the data were previously published in Ref. \citep{pelc2019}.
An extension of the model presented in Sec. \ref{sec:Phenomenology}
that includes the role of inhomogeneity is also introduced. Our conclusions
are presented in Sec. \ref{sec:Conclusions}.

\section{Phenomenological model for the third-harmonic magnetic response \label{sec:Phenomenology}}

In this section, we derive an expression for the third-harmonic magnetic
response $\overline{M_{3}}$, measured in the experiments of Ref.
\citep{pelc2019}, based on the Lawrence-Doniach (LD) approach. We
first review the contribution of the superconducting fluctuations
to the magnetization in the presence of a static magnetic field within
the LD approach. Here we only quote the LD results, which are well
known from the literature -- for their derivations, see for instance
Refs. \citep{larkin2005book,mishonov2000}. Using the LD results,
we then proceed to include an ac field to explicitly calculate $\overline{M_{3}}$,
and discuss its temperature dependence in different regimes.

\subsection{Linear and nonlinear susceptibilities in the Lawrence-Doniach model}

Fluctuations of a superconductor in the presence of an external magnetic
field can be modeled within the phenomenological Ginzburg-Landau framework.
In a regime close to $T_{c}$, the general superconducting Ginzburg-Landau
free-energy functional takes the form: 
\begin{equation}
\begin{aligned}\Delta\mathcal{F}\left[\Psi\left(\mathbf{x}\right)\right]&=  \int d^{d}x\left(a\left|\Psi\right|^{2}+\frac{b}{2}\left|\Psi\right|^{4}\right.\\
 & \left.+\frac{1}{2m^{*}}\left|\left(\frac{\nabla}{i}-e^{*}\mathbf{A}\right)\Psi\right|^{2}+\frac{1}{8\pi}\left|\nabla\times\mathbf{A}\right|^{2}\right)
\end{aligned}
\label{eq_GL}
\end{equation}

Here, $\Psi\left(\mathbf{x}\right)$ is the superconducting order
parameter, $m^{*}=2m$ and $e^{*}=2e$ are the mass and charge of
a Cooper pair, $\mathbf{A}$ is the vector potential, and $b>0$ is
a Ginzburg-Landau parameter. The coefficient $a$ is parametrized
as $a=\alpha\left(T-T_{c}\right)=\alpha T_{c}\epsilon$, where $\epsilon=\frac{T-T_{c}}{T_{c}}$
is the reduced temperature and $\alpha$ a positive constant. Near
$T_{c}$, but above the temperature range where critical fluctuations
become important, as set by the Ginzburg-Levanyuk parameter, one assumes
that the order parameter is small and slowly-varying. As a result,
the quartic term in Eq. (\ref{eq_GL}) can be neglected, and only
Gaussian fluctuations are considered: 
\begin{equation}
\Delta\mathcal{F}\left[\Psi\left(\mathbf{x}\right)\right]=\int{d^{d}x\left(a\left|\Psi\right|^{2}+\frac{1}{4m}\left|\left(\frac{\nabla}{i}-2e\mathbf{A}\right)\Psi\right|^{2}\right)}
\end{equation}

To obtain the Lawrence-Doniach (LD) free-energy expression, one assumes
a layered superconductor and considers a magnetic field $H$ applied
perpendicular to the layers. A detailed derivation can be found in
standard textbooks and review papers, see for instance Refs. \citep{larkin2005book,mishonov2000}.
For completeness, we only highlight the main steps of the derivation
and quote the results from Ref. \citep{larkin2005book}. Because of
the layered nature of the system, there is a difference between in-plane
and out-of-plane kinetic terms. While the former assumes the same
form as in Eq. (\ref{eq_GL}), the latter is described by $\delta_{z}\left|\Psi_{l+1}-\Psi_{l}\right|^{2}$,
where $\delta_{z}$ is the inter-layer coupling constant and the subscript
$l$ is a layer index. It is also convenient to introduce two dimensionless
quantities, $h$ and $r$. By using the result $H_{c2}\left(0\right)=\frac{2m\alpha T_{c}}{e}$
for the zero-temperature critical field, we define the dimensionless
applied field $h\equiv \frac{H}{H_{c2}(0)}$. Moreover, we define the dimensionless
anisotropy parameter $r\equiv\frac{2\delta_{z}}{\alpha T_{c}}$, which
can also be expressed in terms of the ratio between the correlation
length along the $z$ direction, $\xi_{z}(0)$, and the inter-layer
separation $s$, $r=\frac{4\xi_{z}^{2}\left(0\right)}{s^{2}}$. Writing
the order parameter in a product form between in-plane Landau-level
wave functions and plane waves propagating along the $z$ direction,
one can evaluate the partition function $Z=\int{D\Psi D\Psi^{*}e^{-\frac{\Delta\mathcal{F}\left[\Psi\left(\mathbf{x}\right)\right]}{T}}}$
and then obtain the LD free-energy expression (up to a constant) \citep{larkin2005book,mishonov2000}: 

\begin{align}
\frac{F\left(\epsilon\right)}{M_{\infty}H_{c2}(0)} &= -\frac{2\left(\epsilon+1\right)h}{\ln2}\left[\left(\epsilon+\frac{r}{2}\right)\frac{\ln h}{2h}-\frac{1}{2}\ln2\pi\right.\nonumber \\
 & \left.+\int_{0}^{\pi/2}\frac{d\phi}{\pi/2}\,\ln\Gamma\left(\frac{1}{2}+\frac{\epsilon+r\sin^{2}\phi}{2h}\right)\right]
\end{align}

Here, $\Gamma(x)$ is the gamma function, the integration over the
variable $\phi$ effectively sums over the layers, $v$ is the volume,
and $M_{\infty}\equiv\frac{T_{c}}{\Phi_{0}s}\frac{\ln2}{2}$ is the
absolute value of the saturation magnetization at $T_{c}$, with $\Phi_{0}$
denoting the flux quantum. Similarly, the LD expression for the magnetization
is given by \citep{larkin2005book,mishonov2000}:

\begin{equation}
\begin{aligned}\frac{M\left(\epsilon\right)}{M_{\infty}}= & -\frac{2\left(\epsilon+1\right)}{\ln2}\int_{0}^{\pi/2}\frac{d\phi}{\pi/2}\left\{ \frac{\epsilon+r\sin^{2}{\phi}}{2h}\times\right.\\
 & \left[\psi\left(\frac{\epsilon+r\sin^{2}{\phi}}{2h}+\frac{1}{2}\right)-1\right]\\
 & \left.-\ln{\Gamma}\left(\frac{\epsilon+r\sin^{2}{\phi}}{2h}+\frac{1}{2}\right)+\frac{1}{2}\ln{2\pi}\right\} 
\end{aligned}
\label{mag}
\end{equation}
where $\psi(x)=\frac{d\ln\Gamma(x)}{dx}$ is the digamma function.
By taking $h\gg\epsilon,r$ in Eq. (\ref{mag}), the right-hand side
gives $-1$ at $\epsilon=0$, confirming that $M_{\infty}$ is the
saturation magnetization at $T_{c}$. Note that this expression is
valid for $h>0$; in the case of $h<0$, symmetry implies $F\left(-h\right)=F\left(h\right)$
and $M\left(-h\right)=-M\left(h\right)$. For future reference, we
list the three dimensionless parameters that will be employed throughout
this work: 
\begin{equation}
\begin{aligned} & \epsilon=\frac{T-T_{c}}{T_{c}}\\
 & r=\left[\frac{2\xi_{z}\left(0\right)}{s}\right]^{2}\\
 & h=\frac{H}{H_{c2}\left(0\right)}
\end{aligned}
\end{equation}

While the anisotropy parameter $r$ is fixed, its impact on the magnetization
depends on the temperature range probed. In a regime sufficently far
from $T_{c}$, $r\ll\epsilon$, the system essentially behaves as
decoupled layers ($r\rightarrow0$) and Eq.(\ref{mag}) becomes \citep{larkin2005book,mishonov2000}
\begin{equation}
\begin{aligned}\frac{M\left(\epsilon\gg r\right)}{M_{\infty}}= & -\frac{2\left(\epsilon+1\right)}{\ln2}\left\{ \frac{\epsilon}{2h}\left[\psi\left(\frac{1}{2}+\frac{\epsilon}{2h}\right)-1\right]\right.\\
 & \left.-\ln\frac{{\Gamma}\left(\frac{1}{2}+\frac{\epsilon}{2h}\right)}{\sqrt{2\pi}}\right\} .
\end{aligned}
\label{2dmag}
\end{equation}

On the other hand, as $T_{c}$ is approached, the system will eventually
cross over to the regime $r\gg\epsilon$. Then, the three-dimensional
nature of the system cannot be neglected, and the magnetization becomes
\citep{larkin2005book,mishonov2000,kurkijarvi1972}: 
\begin{equation}
\begin{aligned}\frac{M\left(\epsilon\ll r\right)}{M_{\infty}}= & -\frac{6\left(\epsilon+1\right)}{\ln2}\left(\frac{2}{r}\right)^{1/2}\sqrt{h}\left[\zeta\left(-\frac{1}{2},\frac{1}{2}+\frac{\epsilon}{2h}\right)\right.\\
 & \left.-\frac{1}{3}\zeta\left(\frac{1}{2},\frac{1}{2}+\frac{\epsilon}{2h}\right)\frac{\epsilon}{2h}\right]
\end{aligned}
\label{3dmag}
\end{equation}
where $\zeta(\nu,x)$ is Hurwitz zeta function.

\begin{figure}
\centering\includegraphics[width=0.5\textwidth]{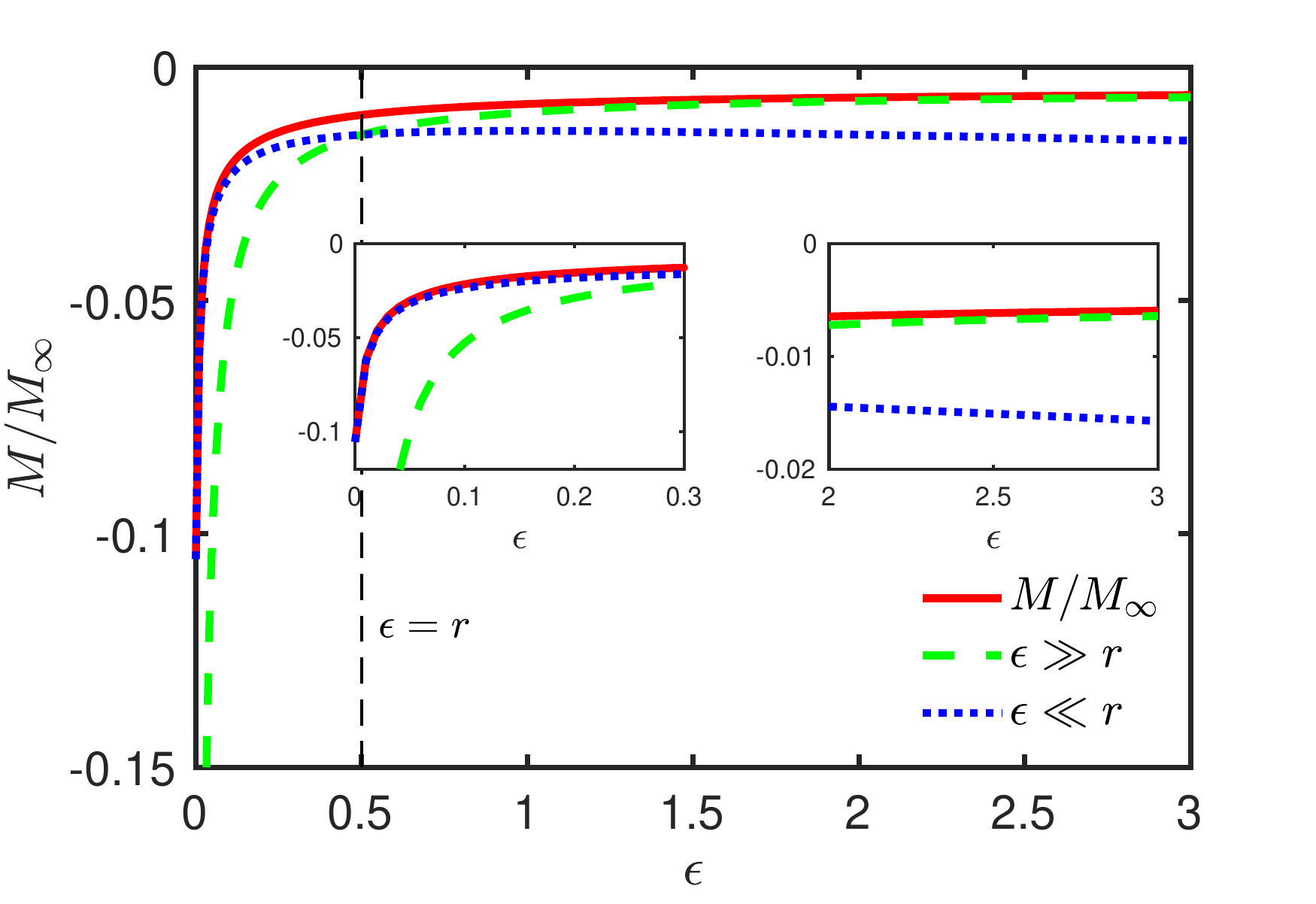} \caption{Magnetization (red curve, in units of $M_{\infty}$) induced by superconducting
fluctuations, in the presence of a dc field $h$, as a function of
the reduced temperature $\epsilon$ according to Eq. (\ref{mag}).
We also include for comparison the asymptotic expressions for $M(\epsilon\gg r)$
(green dashed curve) and $M(\epsilon\ll r)$ (blue dotted curve),
Eqs. (\ref{2dmag}) and (\ref{3dmag}), respectively. A crossover
clearly takes place when $\epsilon\sim r$. The dimensionless parameters
chosen here were $h=0.01,\,r=0.5$. The insets are zooms on different
temperature ranges. }
\label{fig:flucmag} 
\end{figure}

Therefore, as $T_{c}$ is approached from above, we expect a crossover
of the temperature-dependent magnetization from 2D-like behavior to
3D-like behavior, with the crossover temperature corresponding to
$\epsilon\sim r$. This general behavior is illustrated in Fig. \ref{fig:flucmag},
where $M$ given by Eq. (\ref{mag}) is plotted as a function of the
reduced temperature $\epsilon$ together with the asymptotic expressions
in Eqs. (\ref{2dmag})-(\ref{3dmag}) for a fixed field value. As
expected, the contribution of the superconducting fluctuations to
the magnetization are negative.

It will be useful later to contrast the temperature dependence of
the third-harmonic response $\overline{M_{3}}$ with that of the nonlinear
magnetic susceptibility. To derive the latter, we consider the limit
of small fields, \textit{i.e.} when the dimensionless magnetic field
is the smallest parameter of the problem, $h\ll\epsilon,r$. Going
back to the main expression for the magnetization in Eq. (\ref{mag}),
it is convenient to define $y=\frac{\epsilon+r\sin^{2}\phi}{2h}$.
Since $h\ll\epsilon,r$, it follows that $y\gg1$ and the integrand
can be expanded as:

\begin{align}
y\left[\psi\left(y+\frac{1}{2}\right)-1\right]-\ln\Gamma\left(y+\frac{1}{2}\right)+\frac{1}{2}\ln(2\pi) & =\nonumber \\
\frac{1}{12y}-\frac{7}{720y^{3}}+\frac{31}{6720y^{5}}+\mathcal{O}\left(x^{-7}\right)
\end{align}

The integrals over $\phi$ can be analytically evaluated. Expanding
the magnetization in odd powers of $h$,

\begin{equation}
\frac{M}{M_{\infty}}=\chi_{1}h+\chi_{3}h^{3}+\chi_{5}h^{5}+\mathcal{O}(h^{7})\label{M_suscept}
\end{equation}
we find the following expressions for the linear and nonlinear susceptibilities
(see also Refs. \citep{tsuzuki1969,mishonov2000}):

\begin{align}
\chi_{1} & =-\frac{\left(1+\epsilon\right)}{3\ln2}\frac{1}{\epsilon^{1/2}\sqrt{\epsilon+r}}\\
\chi_{3} & =\frac{7\left(1+\epsilon\right)}{360\ln2}\frac{\left(3r^{2}+8r\epsilon+8\epsilon^{2}\right)}{\epsilon^{5/2}\left(\epsilon+r\right)^{5/2}}\label{chi3}\\
\chi_{5} & =-\frac{31\left(1+\epsilon\right)}{13440\ln2}\times\nonumber \\
 & \frac{\left(35r^{4}+160r^{3}\epsilon+288r^{2}\epsilon^{2}+256r\epsilon^{3}+128\epsilon^{4}\right)}{\epsilon^{9/2}\left(\epsilon+r\right)^{9/2}}
\end{align}

Close enough to $T_{c}$, when $\epsilon\ll r$, we find the following
power-law behaviors

\begin{align}
\chi_{1} & \sim-\frac{\epsilon^{-1/2}}{\sqrt{r}}\\
\chi_{3} & \sim\frac{\epsilon^{-5/2}}{\sqrt{r}}\label{chi3_power_law}\\
\chi_{5} & \sim-\frac{\epsilon^{-9/2}}{\sqrt{r}}
\end{align}

\subsection{The third-harmonic magnetic response $\overline{M_3}$: experimental setup and theory}

One of the most common experimental probes of superconducting fluctuations
is to apply a dc magnetic field and measure the magnetic response,
see Eq. (\ref{M_suscept}). The key issue with measuring the linear
susceptibility $\chi_{1}$ is that the diamagnetic contribution due
to the superconducting fluctuations is typically much smaller than
the paramagnetic contributions from other normal-state degrees of
freedom. For the nonlinear susceptibility $\chi_{3}$, however, one
generally expects that the intrinsic normal-state contribution is
negligible in most cases, which could in principle allow one to assess
the contribution from the superconducting fluctuations in a more unambiguous
fashion. Note that, while in principle the susceptibilities $\chi_{1}$
and $\chi_{3}$ are tensor quantities, our experimental setup is designed
in such a way that both the excitation and detection coils are along
the same axis. We therefore only measure in-plane diagonal components,
which are equivalent for a tetragonal or cubic system. Hereafter we
refer only to a scalar $\chi_{3}$.

Instead of applying a dc magnetic field, the experimental technique
presented in Ref. \citep{pelc2019} and utilized here employs an ac
field (of the form $H_{0}\cos{\omega}t$) and a system of coils to
measure the oscillating sample magnetization. In order to determine
the third-order response, a lock-in amplifier is used at the third
harmonic of the fundamental frequency $\omega$, which is typically
in the kHz range. If the fifth-order susceptibility is significantly
smaller than the third-order susceptibility, the third harmonic response
is a good measure of the third-order susceptibility. This condition
was experimentally verified by measuring at the fifth harmonic, where
the signal was found to be vanishingly small except extremely close
to $T_{c}$, where it was still an order of magnitude smaller than
the third harmonic. We can thus safely ignore the higher-order contributions.
Most of the data presented here were published in Ref. \citep{pelc2019},
and were obtained in two separate experimental setups. Low-temperature
measurements on strontium ruthenate were performed in a $^{3}$He
evaporation refrigerator with a custom-made set of coils. Samples
of conventional superconductors were measured in a modified Quantum
Design MPMS, where we used the built-in AC susceptibility coil to
generate the excitation magnetic field, and a custom-made probe with
small detection coils to maximize the filling factor. We estimate
that the magnetization sensitivity of both setups is better than 1
nanoemu, an improvement of 1-2 orders of magnitude over standard SQUID-based
instruments. This is made possible by lock-in detection, matching
the impedance of the detection coils and lock-in amplifier inputs,
and large filling factors of the detection coils \citep{Drobac2013}.

\begin{figure}
\centering\includegraphics[width=0.5\textwidth]{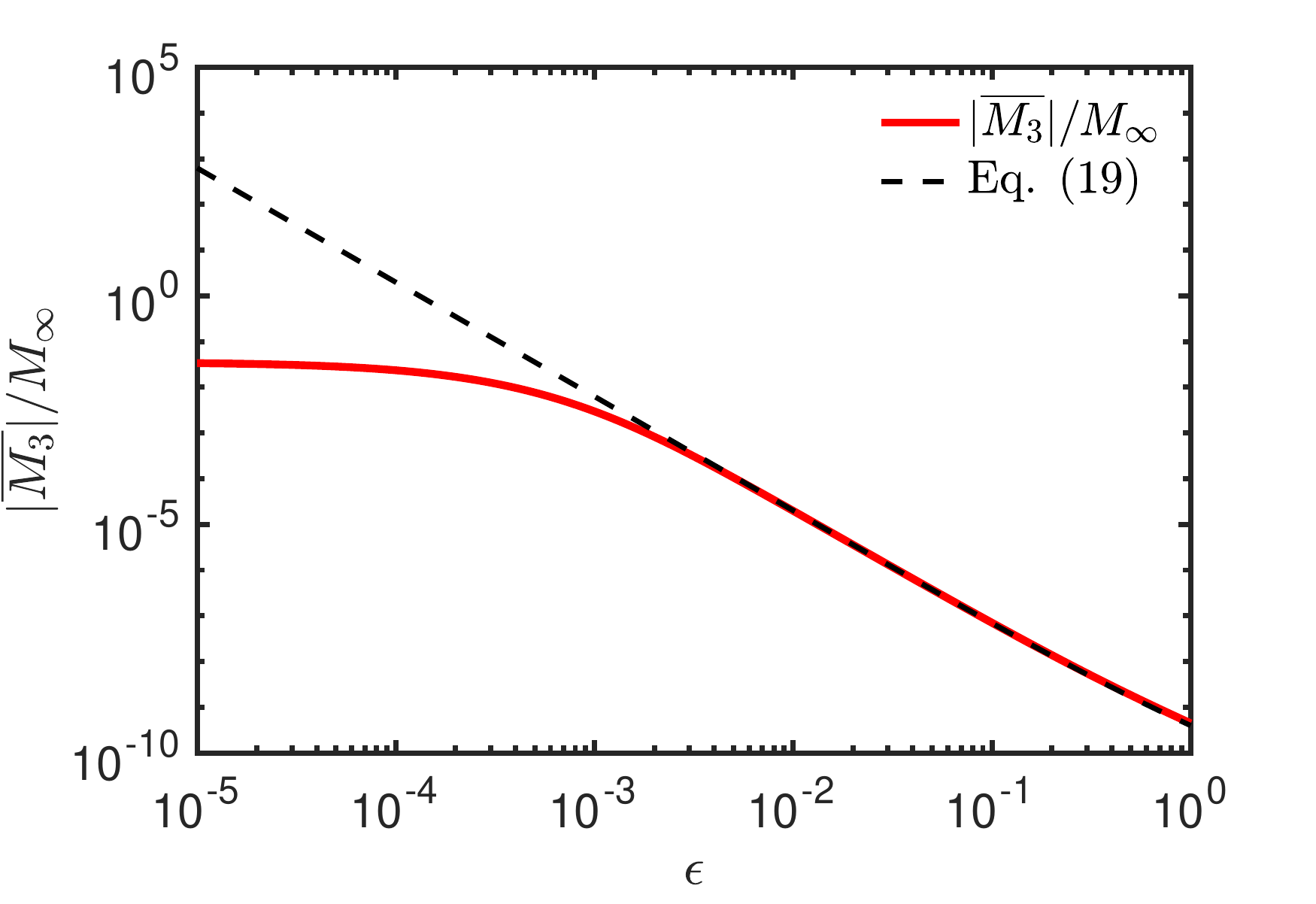} \caption{Absolute value of the third-harmonic response, $|\overline{M_{3}}|$
in Eq. (\ref{numeric}), in units of $M_{\infty}$, as a function
of the reduced temperature $\epsilon\equiv\frac{T-T_{c}}{T_{c}}$,
plotted on a log-log scale (red curve). The dashed black line corresponds
to the analytical approximation in Eq. (\ref{M3_analytics}), which
gives a $\epsilon^{-5/2}$ power-law behavior. The dimensionless parameters
used here are $h_{0}=10^{-3}$ and $r=1$.}
\label{fig:I3log} 
\end{figure}

Although we expect the third-harmonic response to exhibit
behavior similar to the third-order nonlinear susceptibility $\chi_{3}$,
there are important differences, since the amplitude of the oscillating
field, albeit small ($H_{0}\sim1$ Oe), is nonzero. Thus, to provide
a more direct comparison between the LD model and experiments, we
directly compute the third-harmonic response, which we denote by $\overline{M_{3}}$.
In our experimental setup, the signal corresponds to the Fourier transform
of $\frac{\partial M}{\partial t}$ at $3\omega$, 
\begin{equation}
\overline{M_{3}}\left(\epsilon\right)=\int_{-\frac{\pi}{\omega}}^{\frac{\pi}{\omega}}\frac{\partial M\left(\epsilon,h(t)\right)}{\partial t}\,e^{3i\omega t}dt,\label{response}
\end{equation}
where $M\left(\epsilon,h(t)\right)$ is obtained from Eq.(\ref{mag})
by substituting $h=h_{0}\cos{\omega}t$. Integration by parts gives
$\overline{M_{3}}\left(\epsilon\right)=-3i\int_{-\pi}^{\pi}M(\epsilon,h_{0}\cos\theta)e^{3i\theta}d\theta$
with $\theta=\omega t$. Using the fact that $M\left(\epsilon,-h\right)=-M\left(\epsilon,h\right)$,
we have $\int_{-\pi}^{-\pi/2}M(\epsilon,h_{0}\cos\theta)e^{i3\theta}d\theta=\int_{0}^{\pi/2}M(\epsilon,h_{0}\cos\theta)e^{i3\theta}d\theta$
and $\int_{\pi/2}^{\pi}M(\epsilon,h_{0}\cos\theta)e^{i3\theta}d\theta=\int_{-\pi/2}^{0}M(\epsilon,h_{0}\cos\theta)e^{i3\theta}d\theta$,
which yields 
\begin{equation}
\overline{M_{3}}\left(\epsilon\right)=-6i\int_{-\pi/2}^{\pi/2}d\theta\,M\left(\epsilon,h_{0}\cos\theta\right)\cos3\theta,\label{numeric}
\end{equation}
where the field $h_{0}\cos\theta$ remains positive between the integration
limits. Experimentally, both the imaginary and real parts can be measured.
However, due to issues with lock-in phase determination in third-harmonic
measurements \citep{Drobac2013}, we simply use the absolute value
of $\overline{M_{3}}$ for comparison between the experimental and
theoretical results.

\begin{figure}
\begin{centering}
\includegraphics[width=0.5\textwidth]{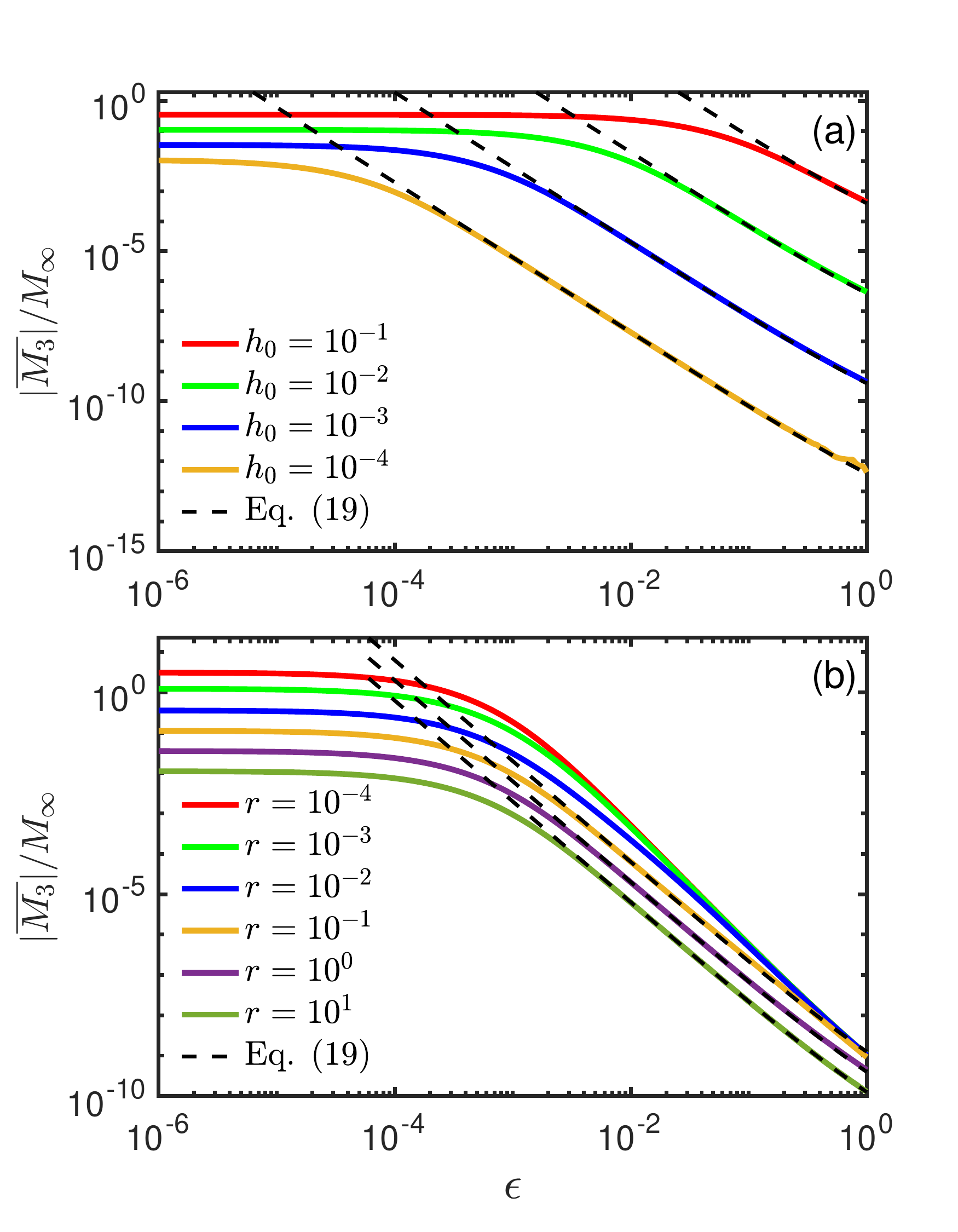} 
\par\end{centering}
\caption{Absolute value of the third-harmonic response $|\overline{M_{3}}|$
(in units of $M_{\infty}$) as a function of the reduced temperature
$\epsilon$ for varying $h_{0}$ values (fixed $r=1$, panel (a))
and varying $r$ values (fixed $h_{0}=10^{-3}$, panel (b)). The dashed
lines mark the power-law behavior $\epsilon^{-5/2}$ displayed by
the curves with larger values of $r$.}
\label{fig:I3varyh} 
\end{figure}

In the temperature range where $h_{0}\ll\epsilon$, we can substitute
the series expansion (\ref{M_suscept}) in Eq. (\ref{numeric}) and
find:

\begin{equation}
\frac{\left|\overline{M_{3}}\right|}{M_{\infty}}\approx\frac{3\pi}{4}\chi_{3}h_{0}^{3}+\frac{15\pi}{16}\chi_{5}h_{0}^{5}.
\end{equation}

Now, in the relevant regime $r\gg\epsilon$, according to Eqs. (\ref{chi3_power_law}),
we have $\chi_{3}\sim\epsilon^{-5/2}$ and $\chi_{5}\sim\epsilon^{-9/2}$.
Therefore, as long as we remain in the regime $h_{0}\ll\epsilon$,
the contribution from the fifth-order nonlinear susceptibility $\chi_{5}$
can be neglected. Using Eq. (\ref{chi3}) we obtain:

\begin{equation}
\frac{\left|\overline{M_{3}}\right|}{M_{\infty}}\approx\left(\frac{7\pi}{160\ln2}\right)\frac{h_{0}^{3}\left(1+\epsilon\right)\epsilon^{-5/2}}{\sqrt{r}}\label{M3_analytics}
\end{equation}

Therefore, we expect that, in the temperature range $h_{0}\ll\epsilon\ll r$,
the third-harmonic response $\left|\overline{M_{3}}\right|$ displays
the power-law behavior $\left(T-T_{c}\right)^{-5/2}$ characteristic
of the third-order nonlinear susceptibility $\chi_{3}$. To verify
this behavior explicitly, in Fig. \ref{fig:I3log} we present the
numerically calculated $|\overline{M_{3}}|$ for $h_{0}=10^{-3}$
and $r=1$, and compare it with the analytical approximation in
Eq. (\ref{M3_analytics}). It is clear that the expected power-law
behavior appears over a rather wide temperature range. As one approaches
$T_{c}$ from above and reaches the temperature scale $\epsilon\sim h_{0}$,
deviations from the power-law are observed, and $\left|\overline{M_{3}}\right|$
saturates to a constant value. This is a direct consequence of the
fact that we are not computing the dc susceptibility, but the ac third-harmonic
response at a fixed field amplitude $h_{0}$. Figs. \ref{fig:I3varyh}(a)-(b)
depict how the temperature window in which power-law behavior is observed
is affected by changing $r$ and $h_{0}$. As expected, increasing
$h_{0}$ significantly suppresses the window of power-law behavior,
as the temperature scale $\epsilon\sim h_{0}$ is moved up. On the
other hand, the anisotropy parameter $r$ has a rather minor impact
on the temperature range in which $\epsilon^{-5/2}$ behavior is observed.

\section{Comparison with experimental data \label{sec:Comparison}}

\subsection{Conventional Superconductors (Pb, Nb, and V)}

In order to validate the LD approach for the third-harmonic response,
we first compare the theoretical results for $\overline{M_{3}}$ from
Eq.(\ref{numeric}) with the experimental third-harmonic data for
three conventional elemental superconductors: lead (Pb), niobium (Nb),
and vanadium (V). Besides an overall pre-factor, there are three fitting
parameters in our formalism: the upper critical field $H_{c2}$, the
critical temperature $T_{c}$, and the anisotropy ratio $r$. The
field $H_{0}$ is 1.3~Oe as generated by the excitation coil, but
the true value could be modified by demagnetization factors (especially
very close and below $T_{c}$) by up to a factor of $\sim2$. Hereafter,
for concreteness, we will use $H_{0}=1.3$ Oe for all cases. Since
these materials are rather three-dimensional, we expect the $z$-axis
correlation length $\xi_{z}$ to be larger than the layer distance
$s$ in the LD model, \textit{i.e.} $r>4$. Thus, because the reduced
temperatures probed are very small ($\epsilon_{\mathrm{max}}\sim10^{-2}$),
the precise value of $r$ does not significantly affect the temperature
dependence of $|\overline{M_{3}}|$ in the experimentally relevant
temperature regime (as shown above in Fig. \ref{fig:I3varyh}(b)). Therefore, to minimize the number of fitting parameters,
we set $r=10$ in all cases. This leaves only two free parameters,
$H_{c2}$ and $T_{c}$.

\begin{figure}
\centering \includegraphics[width=0.5\textwidth]{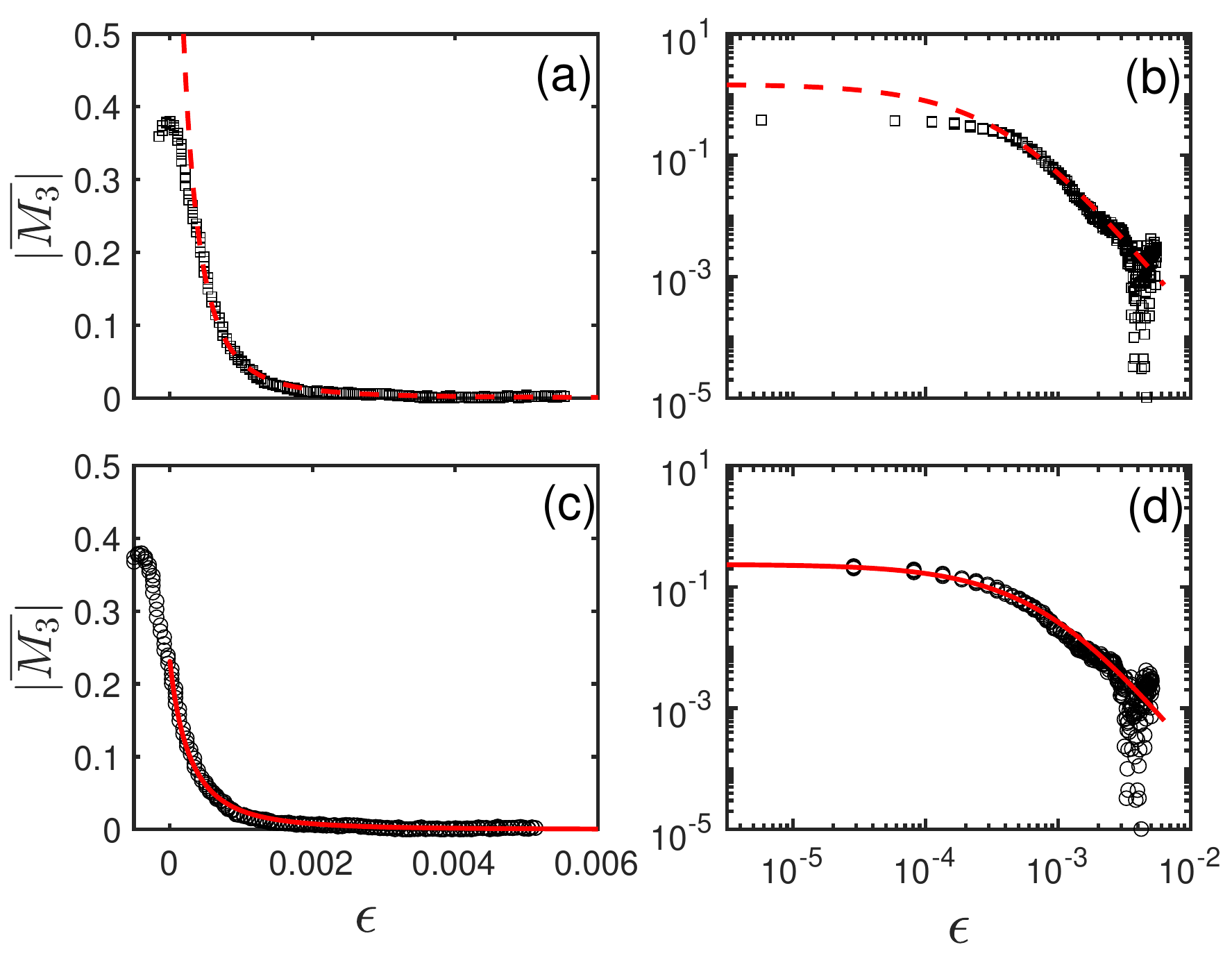} \caption{Comparison between the measured third-harmonic response $\left|\overline{M_{3}}\right|$
(circle and square black symbols, in arbitrary units) for Pb and the
theoretical results obtained from Eq. (\ref{numeric}) (dashed and
solid red lines). Panels (a) and (c) ((b) and (d)) show the data on
a linear (logarithmic) scale. Fit parameters are shown in Table \ref{Table_fitting}.
In panels (a)-(b), the fit parameter is the critical field $\tilde{H}_{c2}$
in Table \ref{Table_fitting}, whereas the critical temperature is
set to its experimental value $T_{c}^{(\mathrm{exp})}$. In panels
(c)-(d), the fit parameters are $H_{c2}$ and $T_{c}$. The anisotropy
parameter is set to $r=10$.}
\label{fig:I3Pb} 
\end{figure}

\begin{figure}
\centering\includegraphics[width=0.5\textwidth]{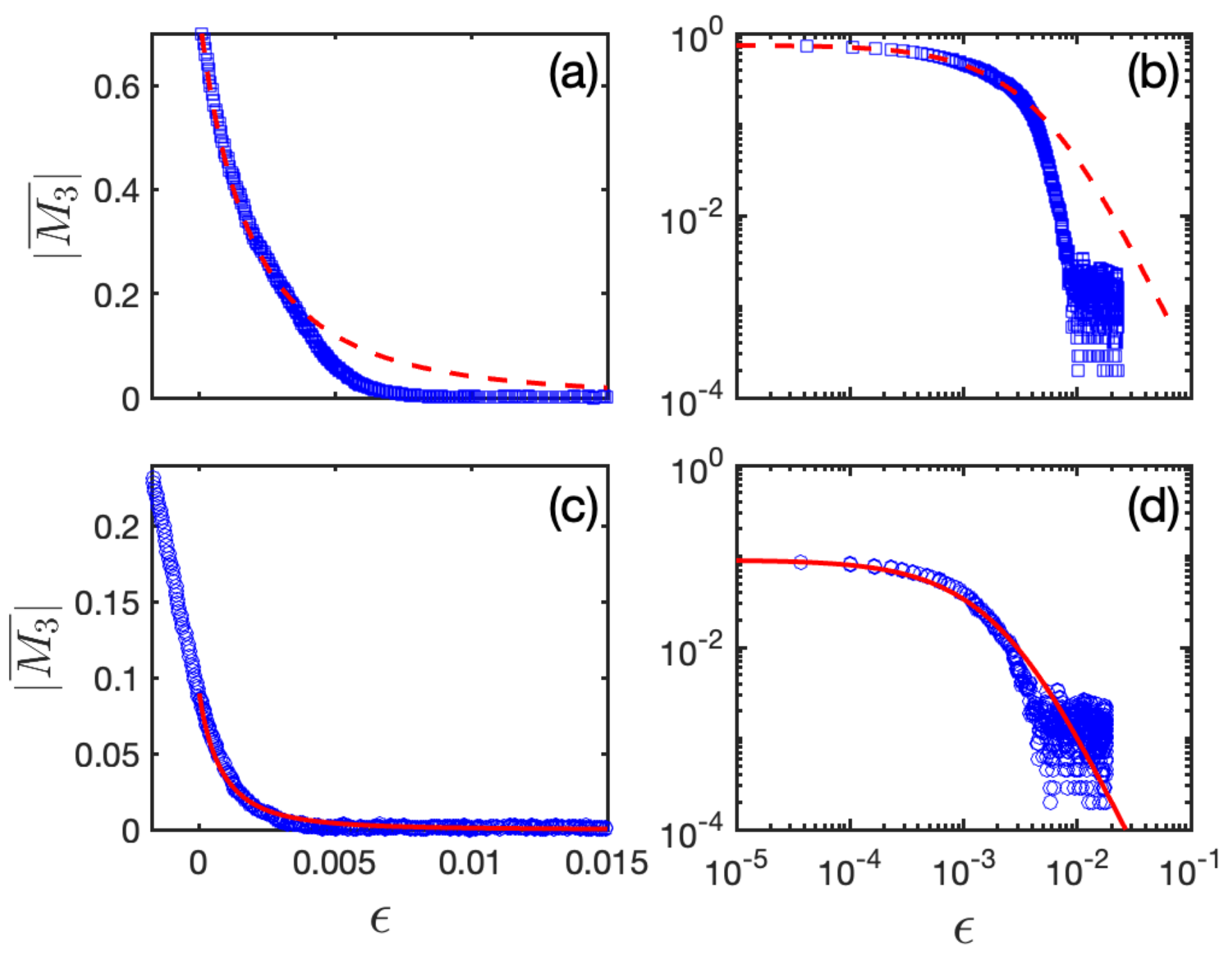} \caption{Comparison between the measured third-harmonic response $\left|\overline{M_{3}}\right|$
(circle and square blue symbols, in arbitrary units) for Nb and the
theoretical results obtained from Eq. (\ref{numeric}) (dashed and
solid red lines). Panels (a) and (c) ((b) and (d)) show the data in
linear (logarithmic) scale. Fit parameters are shown in Table \ref{Table_fitting}.
In panels (a)-(b), the fit parameter is the critical field $\tilde{H}_{c2}$
in Table \ref{Table_fitting}, whereas the critical temperature is
set to its experimental value $T_{c}^{(\mathrm{exp})}$. In panels
(c)-(d), the fitting parameters are $H_{c2}$ and $T_{c}$. The anisotropy
parameter is set to $r=10$. }
\label{fig:I3Nb} 
\end{figure}

\begin{figure}
\centering \includegraphics[width=0.5\textwidth]{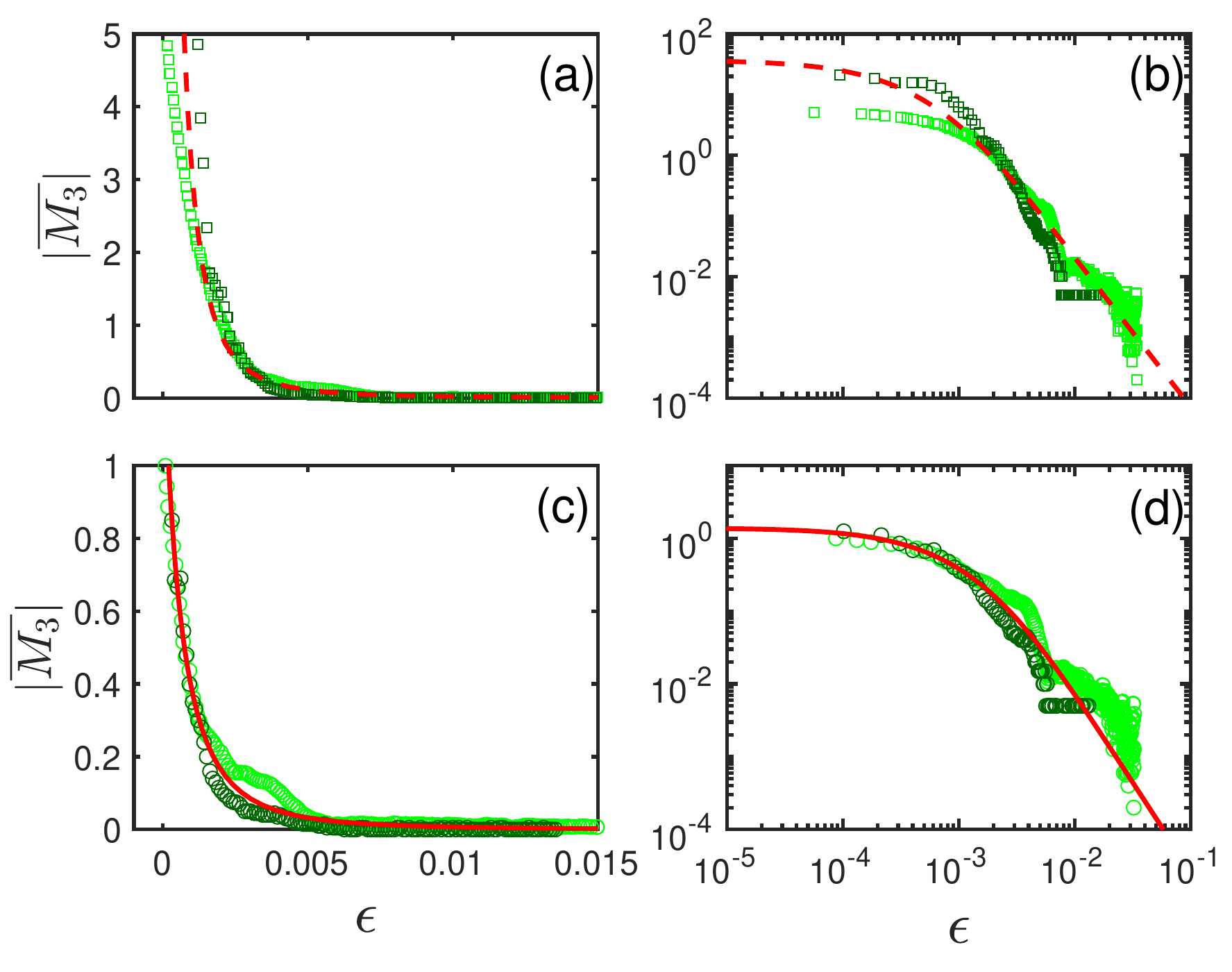} \caption{Comparison between the measured third-harmonic response $\left|\overline{M_{3}}\right|$
(circle and square green symbols, in arbitrary units) for V and the
theoretical results obtained from Eq. (\ref{numeric}) (dashed and
solid red lines). Data from two different samples are presented (light
green and dark green symbols). Panels (a) and (c) ((b) and (d)) show
the data in linear (logarithmic) scale. Fit parameters are shown in
Table \ref{Table_fitting}. In panels (a)-(b), the fit parameter is
the critical field $\tilde{H}_{c2}$ in Table \ref{Table_fitting},
whereas the critical temperature is set to its experimental value
$T_{c}^{(\mathrm{exp})}$. In panels (c)-(d), the fit parameters are
$H_{c2}$ and $T_{c}$. The anisotropy parameter is set to $r=10$.}
\label{fig:I3V} 
\end{figure}

The comparison between theoretical and experimental results is shown
Figs. \ref{fig:I3Pb}, \ref{fig:I3Nb}, and \ref{fig:I3V} for Pb,
Nb, and V, respectively. In all figures, the circle and square symbols
correspond to data, whereas dashed and solid lines correspond to theoretical
results. Experimental measurements of $\left|\overline{M_{3}}\right|$
become challenging below $\epsilon\sim10^{-4}$ due to thermometry
resolution issues, and the signal typically decays below the noise
level around $\epsilon\sim10^{-2}$, indicating a small temperature
regime of significant superconducting fluctuations. In the case of
V, a kink is observed in one sample (light green symbols), which is
possibly a spurious signal due to solder superconductivity or the
result of a slight macroscopic sample inhomogeneity. For this reason,
we also include results from a second sample (dark green symbols).
Because the overall magnitude of the experimental $\left|\overline{M_{3}}\right|$
is arbitrary and changes with modifications of the set-up, we rescaled
the $\left|\overline{M_{3}}\right|$ values of the second sample (dark
green symbols) by an overall constant to better match the behavior
of $\left|\overline{M_{3}}\right|$ of the first sample (light green
symbols) at larger $\epsilon$ values. 

In order to obtain the best
fit, we considered two slightly different procedures. In panels (a)-(b)
of each figure (dashed lines), we fixed $T_{c}$ to be the temperature
at which the third-harmonic response displays a maximum; we refer
to this value as $T_{c}^{(\mathrm{exp})}$. It is important to note,
however, that this value is not necessarily the exact temperature
of zero resistance onset. For this reason, and given the intrinsic
experimental uncertainties in the precise absolute determination of
$T_{c}$, in panels (c)-(d) (solid lines) we allowed $T_{c}$ to vary
from $T_{c}^{(\mathrm{exp})}$, but by no more than $0.5\%$. The
fit parameters are shown in Table \ref{Table_fitting}, together with
the experimental values for $T_{c}^{(\mathrm{exp})}$ and $H_{c2}^{(\mathrm{exp})}$,
the latter taken from Ref. \citep{lide2004crc}. Note that, to distinguish
between the two fitting procedures, we denote by $\tilde{H}_{c2}$
the value used in panels (a)-(b) of the figures. Moreover, since Pb
is a type-I superconductor, $H_{c2}^{(\mathrm{exp})}$ was estimated
through $\sqrt{2}\kappa H_{c}$ \citep{tinkham2004book}, with $\kappa=0.24$
\citep{smith1969PbkappaTcHc,Pbkappa1969} and $H_{c}=803$ Oe \citep{smith1969PbkappaTcHc,martienssen2006springer}.

Panels (a)-(b) of Figs. \ref{fig:I3Pb}, \ref{fig:I3Nb}, and \ref{fig:I3V}
show that the theoretical curves obtained by fixing $T_{c}=T_{c}^{(\mathrm{exp})}$
provide a reasonable description of the third-harmonic data in the
region not too close to $T_{c}$ for Pb and V (Figs. \ref{fig:I3Pb}
and \ref{fig:I3V}), and in the region close to $T_{c}$ for Nb (Fig.
\ref{fig:I3Nb}). In particular, the latter does not seem to display
the characteristic $\epsilon^{-5/2}$ power-law behavior observed
in the former two in the regime of intermediate $\epsilon$ values.
However, because of the definition of the reduced temperature, $\epsilon=\frac{T-T_{c}}{T_{c}}$,
even small changes in $T_{c}$ within typical experimental uncertainty
could account for these deviations between theory and experiment.
As noted above, to address this issue we performed a second fit procedure
allowing $T_{c}$ to be slightly different than $T_{c}^{(\mathrm{exp})}$.
As shown in panels (c)-(d) of the same figures, we find a better agreement
between the theoretical and experimental results over a wider temperature
range, including in the case of Nb in the intermediate $\epsilon$
range. Comparing the theoretical $T_{c}$ values in Table \ref{Table_fitting}
with the $T_{c}^{(\mathrm{exp})}$ values, we note that in all cases
$T_{c}$ is slightly larger than $T_{c}^{(\mathrm{exp})}$. This is
the reason why in panels (c)-(d) the theoretical curves stop at $\epsilon=0$
whereas the data extend to the region $\epsilon<0$.

On the other hand, there is a more significant difference between
$H_{c2}$ and the experimental value $H_{c2}^{(\mathrm{exp})}$ taken
from the literature, with the former being a factor of approximately
$2$ to $6$ smaller or larger than the latter. We note that the intrinsic
uncertainty in the precise value of $H_{0}$ in our experiment may
explain at least part of this discrepancy. Moreover, the value of
$H_{c2}^{(\mathrm{exp})}$ strongly depends on material preparation
details, especially for polycrystalline samples where significant
internal strains can be present \citep{van1967NbVTcHc2}. In principle,
the critical fields are lower in more pristine materials, and it is
therefore meaningful to take the lowest known experimental values
(taken from Ref. \citep{lide2004crc}) for our comparison. Finally,
while the LD model employed here to calculate $\left|\overline{M_{3}}\right|$
assumes a layered system, the bulk elemental superconductors are cubic.
On top of that, the LD approach of including only Gaussian fluctuations
is expected to break down below a very small $\epsilon_{\mathrm{crit}}$,
whose precise value is likely different for distinct materials. Despite
these drawbacks, this comparison shows that the LD model for the third-harmonic
response $\left|\overline{M_{3}}\right|$ due to contributions from
superconducting fluctuations provides a satisfactory description of
the experimental results.

\begin{table}
\begin{centering}
\begin{tabular}{|c|c|c|c|c|c|}
\hline 
 & $T_{c}^{(\mathrm{exp})}$(K)  & $H_{c2}^{(\mathrm{exp})}$(G)  & $\tilde{H}_{c2}$ (G)  & $H_{c2}$ (G)  & $T_{c}^{(\mathrm{exp})}/T_{c}$\tabularnewline
\hline 
\hline 
Pb  & 7.18  & 273  & 2170  & 1083  & 0.9996\tabularnewline
\hline 
Nb  & 9.31  & 1710 & 166  & 286  & 0.9955\tabularnewline
\hline 
V  & 5.29  & 1200 & 1300  & 520 & 0.9980\tabularnewline
\hline 
\end{tabular}
\par\end{centering}
\caption{Experimental critical temperature and critical field values, $T_{c}^{(\mathrm{exp})}$
and $H_{c2}^{(\mathrm{exp})}$, compared to the theoretical fitting
parameters $T_{c}$,\textcolor{blue}{{} }$\tilde{H}_{c2}$ and $H_{c2}$.
$\tilde{H}_{c2}$ corresponds to the fits in panels (a)-(b) of Figs.
\ref{fig:I3Pb}, \ref{fig:I3Nb}, and \ref{fig:I3V}, where $T_{c}$
is forced to be equal to the temperature where the experimental third-harmonic
response displays a maximum (denoted here by $T_{c}^{(\mathrm{exp})}$).
On the other hand, $H_{c2}$ corresponds to the fits in panels (c)-(d)
of the same figures, where $T_{c}$ is allowed to be different from
the experimental value. The $H_{c2}^{(\mathrm{exp})}$ values for
Nb and V are the smallest ones reported in Ref. \citep{lide2004crc},
whereas $H_{c2}^{(\mathrm{exp})}$ for Pb was estimated as explained
in the text. \label{Table_fitting}}
\end{table}

\subsection{Strontium Ruthenate (Sr$_{2}$RuO$_{4}$)}

Having validated our theoretical approach to compute the third-harmonic
response $\left|\overline{M_{3}}\right|$ by comparison with data
for elemental superconductors, we now perform the same comparison
with the lamellar perovskite-derived superconductor Sr$_{2}$RuO$_{4}$
(SRO). The main advantage of our LD calculation of $\left|\overline{M_{3}}\right|$
is that it is entirely phenomenological and independent of microscopic
details. In fact, the main assumption is that the superconducting
fluctuations can be described by a Gaussian approximation. Consequently,
the calculation could in principle be applicable to unconventional
superconductors as well.

SRO is believed to host an unconventional superconducting state that
breaks time-reversal symmetry \citep{Luke1998,Kapitulnik2006,Grinenko2021}.
Whereas for a long time SRO was considered a promising candidate for
$p$-wave triplet superconductivity \citep{Mackenzie2003,Kallin2012},
recent experiments have revealed problems with this interpretation
\citep{mackenzie2017,Pustogow2019,Chronister2020}. This has motivated
alternative proposals involving \textit{e.g.} $d$-wave and $g$-wave
superconductivity \citep{Simon2019,Ramires2019,Romer2019,Agterberg2020,Kivelson2020,Willa2020}.
As mentioned above, the data presented here are the same as in Ref.
\citep{pelc2019}. As shown there, the third-harmonic response of
other perovskite-based superconductors like strontium titanate and
the cuprates display a similar unusual temperature dependence.

\begin{figure}
\begin{centering}
\includegraphics[width=0.5\textwidth]{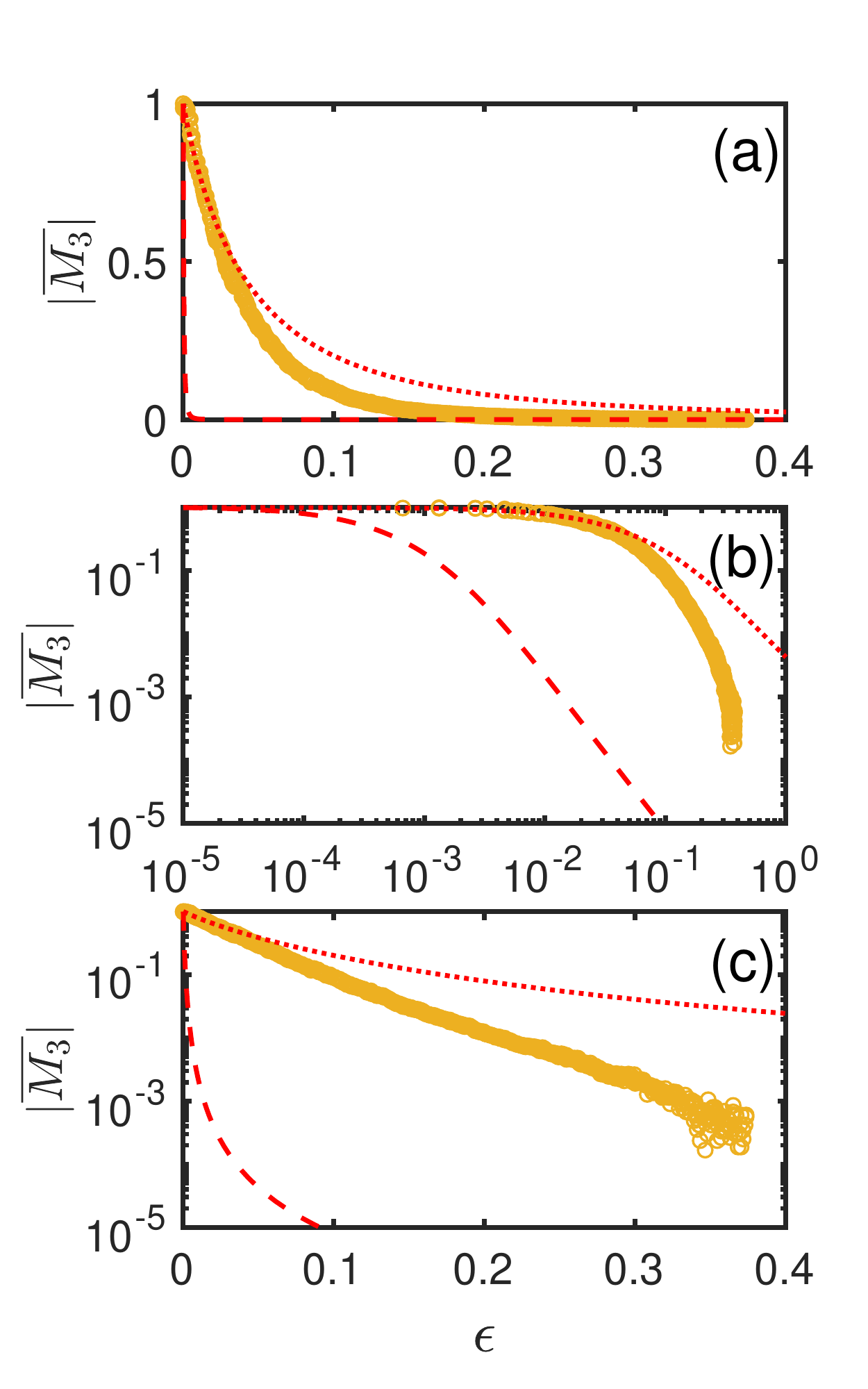}
\par\end{centering}
\caption{Comparison between the experimentally measured third-harmonic response
$\left|\overline{M_{3}}\right|$ (orange symbols, in arbitrary units)
for SRO and the theoretical results obtained from Eq. (\ref{numeric})
(dashed and dotted red lines). Panels (a), (b), and (c) show the data
on linear, logarithmic, and semi-logarithmic scale, respectively.
For the theoretical curves, the critical temperature is set to its
experimental value $T_{c}^{(\mathrm{exp})}$ whereas the critical
field is set to $H_{c2}^{(\mathrm{exp})}$ (dashed lines) and to $0.01H_{c2}^{(\mathrm{exp})}$
(dotted lines). The anisotropy parameter is set to $r=10$.}
\label{fig:I3SRO} 
\end{figure}

The data for SRO are shown by the orange symbols in Fig. \ref{fig:I3SRO}
on linear scale (panel (a)), logarithmic scale (panel (b)), and semi-logarithmic
scale (panel (c)). The theoretical results for $\left|\overline{M_{3}}\right|$
are plotted in the same panels using the experimental critical temperature
value, $T_{c}=1.51\:\mathrm{K}=T_{c}^{(\mathrm{exp})}$, and two different
critical field values: $H_{c2}=750\:\mathrm{G}=H_{c2}^{(\mathrm{exp})}$
(dashed lines) and $H_{c2}=7.6\:\mathrm{G}\approx0.01H_{c2}^{(\mathrm{exp})}$
(dotted lines). Here, $T_{c}^{(\mathrm{exp})}$ corresponds to the
temperature at which the third-harmonic response is maximum, and $H_{c2}^{(\mathrm{exp})}$
is the experimental value reported in the literature \citep{Mackenzie2003,kittaka2009}.
The key observation is that the theoretical $\left|\overline{M_{3}}\right|$
curve with $H_{c2}=H_{c2}^{(\mathrm{exp})}$ grossly underestimates
the data. It is necessary to reduce $H_{c2}$ by two orders of magnitude
to obtain values that are comparable between theory and experiment.
In contrast, for the elemental superconductors, the difference in
the theoretical and experimental $H_{c2}$ values was at most a factor
of $6$. More importantly, even by changing $H_{c2}$ by such a large
amount, the temperature dependence of the data is not captured by
the theoretical $\left|\overline{M_{3}}\right|$ curve, in contrast
again to the case of conventional superconductors. Indeed, while the
theoretical $\left|\overline{M_{3}}\right|$ curve shows a power-law
for intermediate reduced temperatures, the data display an accurately
exponential temperature dependence, as discussed in Ref. \citep{pelc2019}
and shown in panel (c) of Fig. \ref{fig:I3SRO}. We note that the
experimental $H_{c2}$ value depends very strongly on the orientation
of the field with respect to the crystalline $c$-axis, such that
a small misalignment can lead to sizable variation \citep{kittaka2009}.
However, the discrepancy between the theoretical and experimental
results cannot be explained by sample misalignment, since the critical
field \textit{increases} with increasing angle between the field direction
and the crystalline $c$-axis, whereas our theoretical results require
\textit{smaller} $H_{c2}$ values. 

\begin{figure}
\centering \includegraphics[width=0.5\textwidth]{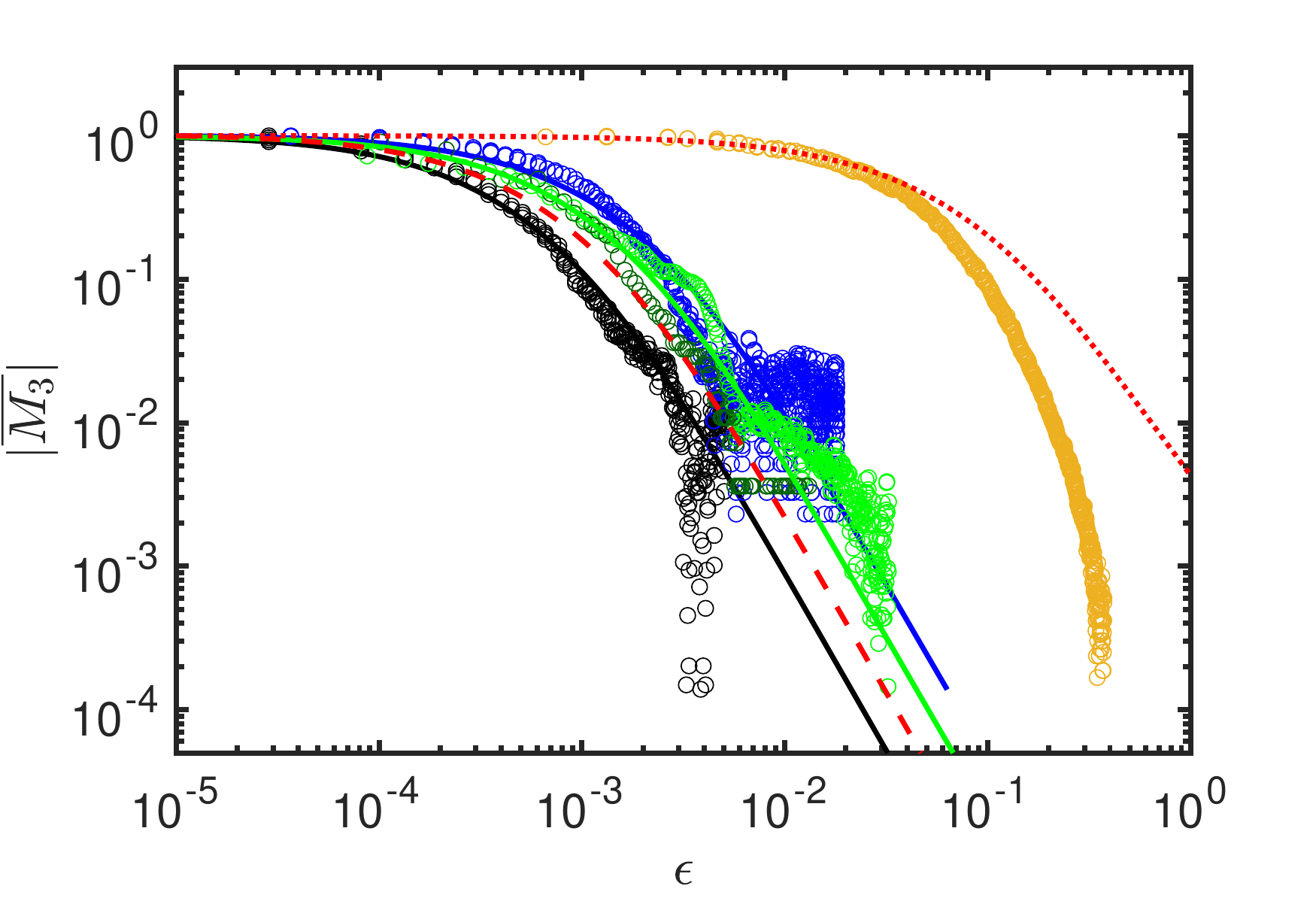} \caption{Comparison between the normalized third-harmonic response data and
the theoretical $\left|\overline{M_{3}}\right|$ results for Pb, Nb,
V and SRO on a logarithmic scale. The solid lines correspond
to the best fits in Figs. \ref{fig:I3Pb}, \ref{fig:I3Nb}, and \ref{fig:I3V},
which refer to the conventional superconductors, whereas the dashed
and dotted lines correspond to the fits for SRO in Fig. \ref{fig:I3SRO}.}
\label{fig:I3all} 
\end{figure}

Fig. \ref{fig:I3all} summarizes the third-harmonic response $\left|\overline{M_{3}}\right|$
of the three conventional superconductors studied here (Pb, Nb, V),
as well as of the unconventional superconductor SRO. The differences
between SRO and the conventional superconductors are not only on the
temperature dependence of $\left|\overline{M_{3}}\right|$, but also
on the fact that $\left|\overline{M_{3}}\right|$ is larger and extends
over a much wider relative temperature range in SRO. Indeed, while
superconducting fluctuations are detected up to $\epsilon\sim10^{-2}$
in conventional superconductors, they extend all the way up to $\epsilon\sim1$
in SRO.

To attempt to address the discrepancy between the theoretical and
experimental results for SRO, we revisit the assumptions behind the
LD model, from which we derived the expression for $\left|\overline{M_{3}}\right|$.
As discussed above, the LD model makes no reference to the microscopic
pairing mechanism. However, it does assume a homogeneous system. In
contrast, perovskites are known for their intrinsic inhomogeneity,
arising from \textit{e.g.} oxygen vacancies and local structural distortions
that deviate strongly from the average lattice structure (see \citep{pelc2021}
and references therein). Indeed, the experiments of Ref. \citep{pelc2019}
indicate that universal structural inhomogeneity is present in perovskite-based
superconductors such as SRO. It has also been argued that dislocations
can have a strong impact on the superconducting state properties of
several perovskites \citep{Ying2013,Hameed2020,Willa2020}. In the
particular case of SRO, muon spin-rotation measurements find a rather
inhomogeneous signature of time-reversal symmetry-breaking below $T_{c}$
\citep{Grinenko2021}. It is also known that the $T_{c}$ of SRO is
strongly dependent on stress \citep{steppke2017,Grinenko2021}, implying
that inhomogeneous internal stresses would lead to regions with locally
modified $T_{c}$. Simple point disorder also leads to a variation
of the local critical temperature \citep{mackenzie1998}. Indeed,
scanning SQUID measurements have directly detected $T_{c}$ inhomogeneity
on the micron scale \citep{watson2018}.

\begin{figure}
\centering \includegraphics[width=0.5\textwidth]{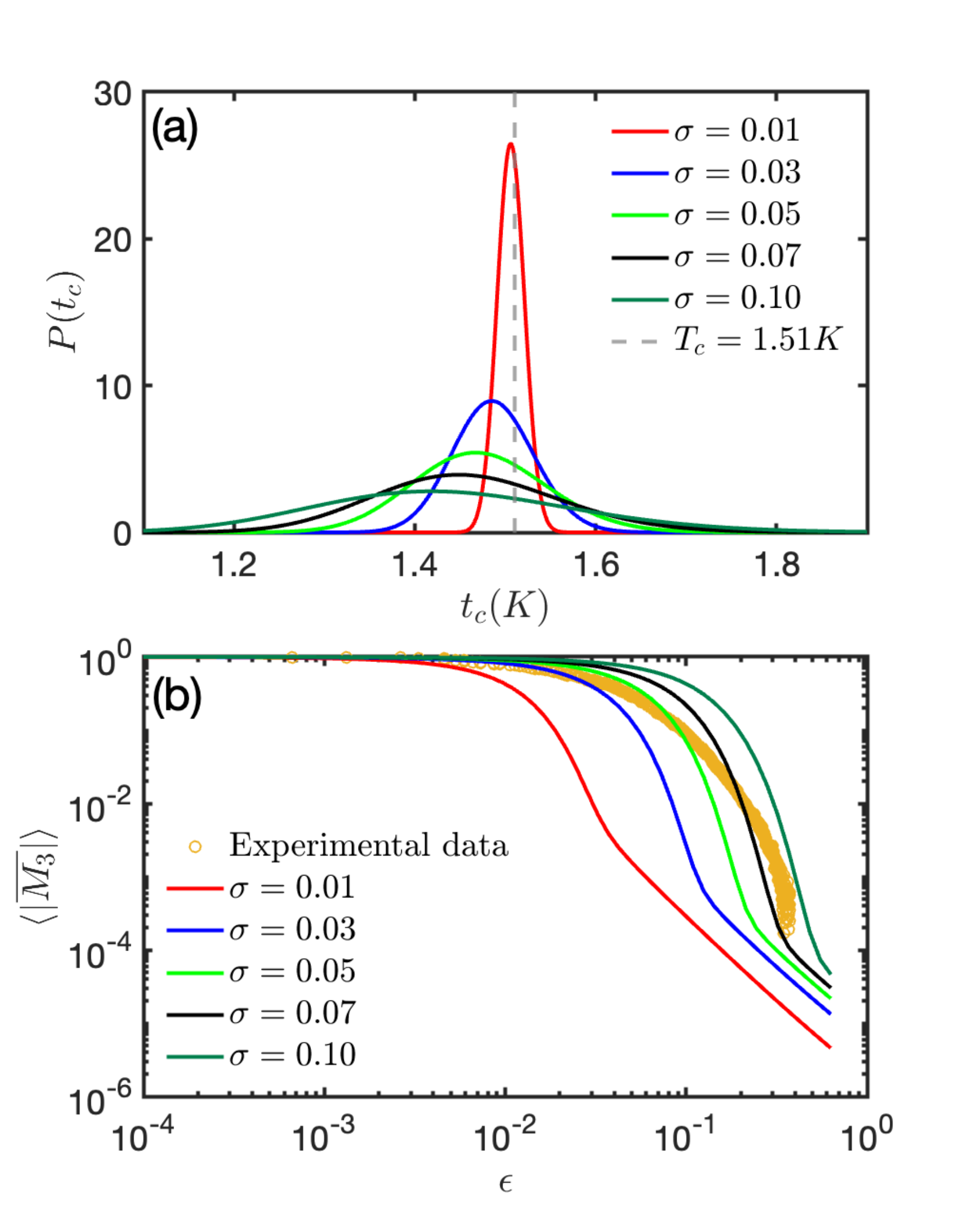} \caption{(a) Normalized probability distribution function of the critical temperature
$t_{c}$ for different values of the parameter $\sigma$ in Eq. (\ref{P_tc}).
Here, the parameter $\mu$ is fixed by the condition $v_{F}\left(T_{c}^{(\mathrm{exp})}\right)=0.3$,
with $T_{c}^{(\mathrm{exp})}=1.51$ K (indicated by the dashed gray
vertical line) and the temperature-dependent superconducting volume
fraction $v_{F}$ defined by Eq. (\ref{volume_fraction}). (b) Averaged
third-harmonic response $\left\langle \left|\overline{M_{3}}\right|\right\rangle $
calculated from the distribution functions of panel (a), compared
to the data for SRO, as a function of $\epsilon=\frac{T}{T_{c}}-1$.
In this calculation, we used the experimental values $T_{c}^{(\mathrm{exp})}=1.51\:\mathrm{K}$
and $H_{c2}^{(\mathrm{exp})}=750\:\mathrm{G}$, and set $r=10$.}
\label{fig:ln} 
\end{figure}

The impact of inhomogeneity on superconducting properties has been
studied by a variety of approaches \citep{coleman1995,andersen2006,Trivedi2014,Pelc2018,dodaro2018}.
Here, we consider a phenomenological approach that introduces a probability
distribution of the local $T_{c}$ (see also Ref. \citep{mayoh2015}).
Such an inhomogeneous $T_{c}$ distribution may explain why the superconducting
fluctuations in SRO are stronger and extend to higher reduced temperatures
as compared to conventional superconductors, since regions with a
locally higher $T_{c}$ are expected to result in a much larger contribution
to $\left|\overline{M_{3}}\right|$ than that arising from the rest
of the sample. To test this idea, we include a distribution function
for $T_{c}$ into our LD-based phenomenological model. We denote the
``transition temperature variable'' as $t_{c}$, and reserve the
notation $T_{c}$ for the actual transition temperature of the system
to avoid confusion. The form of the distribution function $P(t_{c})$
depends on several sources of inhomogeneity in the system, see for
instance Ref. \citep{dodaro2018}. A microscopic derivation is thus
very challenging, and beyond the scope of this work. Instead, here
we opt for a simple phenomenological modeling of $P(t_{c})$. In particular,
we employ a normalized log-normal distribution:

\begin{equation}
P\left(t_{c}\right)=\frac{1}{t_{c}\sqrt{2\pi\sigma^{2}}}\exp\left[-\frac{\left(\ln\frac{t_{c}}{\mu}\right)^{2}}{2\sigma^{2}}\right]\label{P_tc}
\end{equation}
where $\mu$ and $\sigma$ are positive parameters that determine
the mean value and variance of the distribution. The choice of this
distribution is motivated by its properties of only allowing non-zero
values of $t_{c}$ and of having long tails toward larger values of
$t_{c}$. We note that a log-normal distribution for the local gap
-- and consequently of the local $T_{c}$ -- was previously derived
theoretically in Ref. \citep{mayoh2015} for disordered quasi-two-dimensional
superconductors in the limit of weak multifractality, and observed
experimentally in weakly disordered monolayer NbSe$_{2}$ \citep{Rubio2020}.
The averaged fluctuation magnetization in Eq.(\ref{mag}) acquires
the following form:

\begin{equation}
\left\langle M\right\rangle \left(\epsilon\right)=\int_{0}^{T}\,\frac{dt_{c}}{t_{c}\sqrt{2\pi\sigma^{2}}}\exp\left[-\frac{\left(\ln\frac{t_{c}}{\mu}\right)^{2}}{2\sigma^{2}}\right]M\left(\frac{T}{t_{c}}-1\right)
\end{equation}
with $M\left(\epsilon\right)$ given by Eq. (\ref{mag}). We can then
compute the averaged third-harmonic response $\left\langle \left|\overline{M_{3}}\right|\right\rangle $
from Eq. (\ref{numeric}). We assume that $\left\langle \left|\overline{M_{3}}\right|\right\rangle $
is dominated by superconducting fluctuations contributions, which
appear only in regions that are locally non-superconducting (\textit{i.e.}
for which $\epsilon=\frac{T}{t_{c}}-1$ is positive). For this reason,
the limits of the $t_{c}$ integration are such that $0<t_{c}<T$.

\begin{figure}
\centering \includegraphics[width=0.5\textwidth]{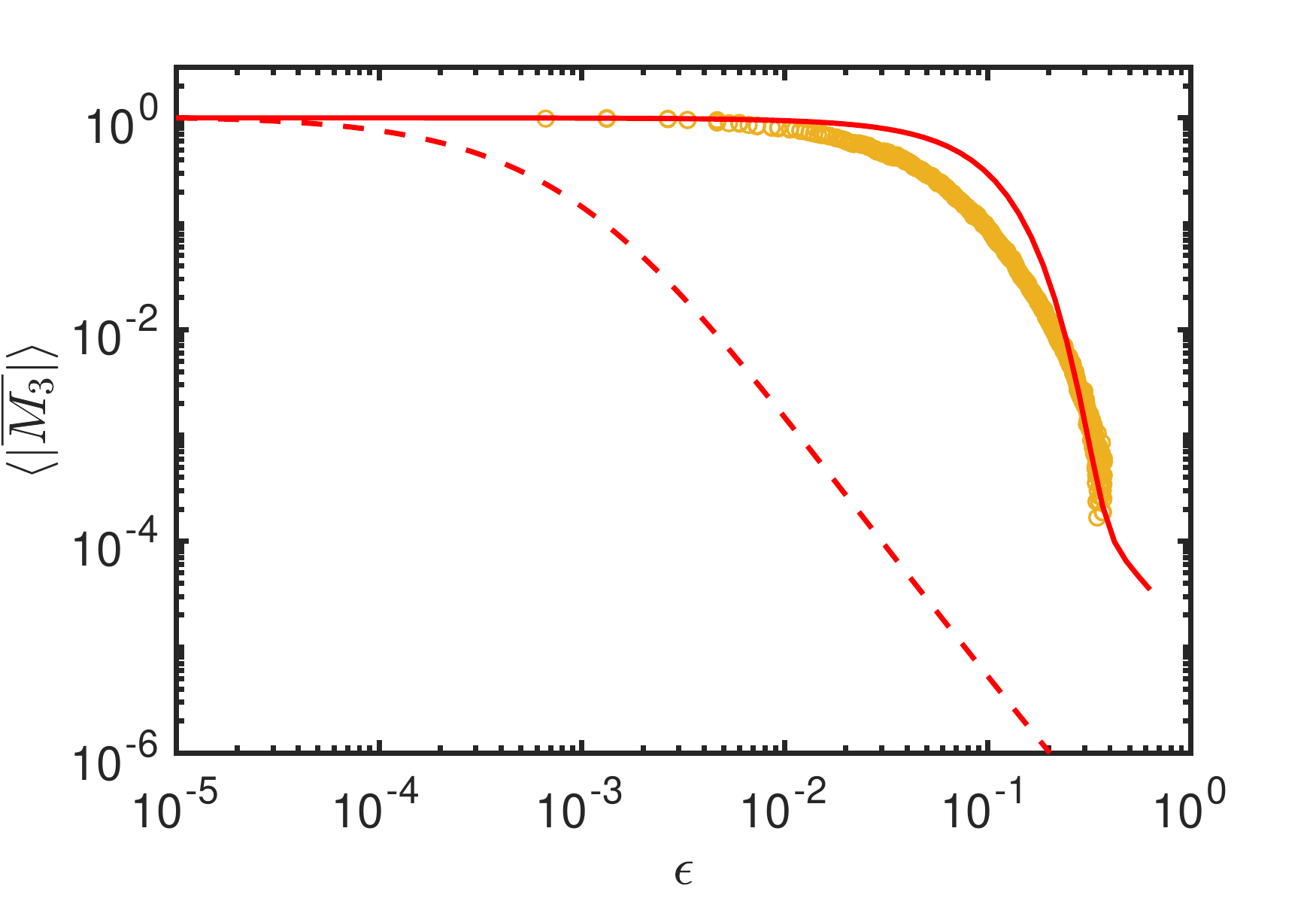} \caption{Averaged third-harmonic response $\left\langle \left|\overline{M_{3}}\right|\right\rangle $
as a function of the reduced temperature $\epsilon=\frac{T}{T_{c}}-1$
calculated using the parameters $T_{c}=1.41\:\mathrm{K}\approx0.93T_{c}^{(\mathrm{exp})}$
and $\sigma=0.08$, while keeping $H_{c2}=H_{c2}^{(\mathrm{exp})}=750\:\mathrm{G}$
and $r=10$ (solid red line). The orange symbols are the experimental
results, and the dashed red line reproduces the theoretical third-harmonic
response $\left|\overline{M_{3}}\right|$ of the clean system with
$T_{c}=T_{c}^{(\mathrm{exp})}$ and $H_{c2}=H_{c2}^{(\mathrm{exp})}$.}
\label{fig:final}
\end{figure}

The two parameters characterizing the distribution function, $\mu$
and $\sigma$, are not independent, since they are related by the
value of $T_{c}$. To see that, we first define the temperature-dependent
superconducting volume fraction $v_{F}\left(T\right)$, which is given
by

\begin{equation}
v_{F}\left(T\right)=1-\int_{0}^{T}{P\left(t_{c}\right)dt_{c}}=\frac{1}{2}-\frac{1}{2}\mathrm{erf}\left(\frac{\ln\frac{t_{c}}{\mu}}{\sqrt{2}\sigma}\right),\label{volume_fraction}
\end{equation}
since the integral on the right-hand side gives the non-superconducting
volume fraction ($T>t_{c}$). When the volume fraction becomes larger
than a threshold value $v_{F}^{*}$, the local superconducting regions
are expected to percolate and the whole sample becomes superconducting.
Note that a similar criterion was used in the analysis of Ref. \citep{mayoh2015}.
$T_{c}$ is then obtained by solving the equation $v_{F}\left(T_{c}\right)=v_{F}^{*}$,

\begin{equation}
\frac{\mu}{T_{c}}=\exp\left[-\sqrt{2}\sigma\mathrm{erf}^{-1}\left(1-2v_{F}^{*}\right)\right],
\end{equation}
where $\mathrm{erf}^{-1}(x)$ is the inverse error function. For simplicity,
we use for $v_{F}^{*}$ the site percolation threshold value for a
cubic lattice, $v_{F}^{*}=0.3$. While $v_{F}^{*}$ itself could be
considered a free parameter, we opt to fix it to avoid increasing
the number of fitting parameters. As a result, the only additional
parameter needed to compute $\left\langle \left|\overline{M_{3}}\right|\right\rangle $,
as compared to the ``clean'' system $\left|\overline{M_{3}}\right|$,
is the dimensionless $\sigma$, which determines the width of the
distribution. In Fig. \ref{fig:ln}(a), we illustrate the profile
of $P\left(t_{c}\right)$ for different values of $\sigma$ under
the constraint $v_{F}\left(T_{c}^{(\mathrm{exp})}\right)=0.3$. The
full expression for $\left\langle \left|\overline{M_{3}}\right|\right\rangle $
then becomes:
\begin{widetext}
\begin{equation}
\frac{\left\langle \left|\overline{M_{3}}\right|\right\rangle \left(\epsilon\right)}{M_{\infty}}=\frac{24\left(\epsilon+1\right)}{\pi\ln2}\int_{0}^{1}\frac{dx}{x\sqrt{2\pi\sigma^{2}}}\exp\left\{ -\left[\frac{\ln\left(x\epsilon+x\right)}{\sqrt{2}\sigma}+\mathrm{erf}^{-1}\left(1-2v_{F}^{*}\right)\right]^{2}\right\} \int_{-\pi/2}^{\pi/2}\mathcal{M}\left(x,h_{0}\cos\theta\right)\cos3\theta\,d\theta
\end{equation}
with:

\begin{equation}
\mathcal{M}\left(x,h\right)=-\int_{0}^{\frac{\pi}{2}}d\phi\left\{ \frac{\frac{1}{x}-1+r\sin^{2}{\phi}}{2h}\left[\psi\left(\frac{\frac{1}{x}-1+r\sin^{2}{\phi}}{2h}+\frac{1}{2}\right)-1\right]-\ln{\Gamma}\left(\frac{\frac{1}{x}-1+r\sin^{2}{\phi}}{2h}+\frac{1}{2}\right)+\frac{1}{2}\ln{\left(2\pi\right)}\right\} 
\end{equation}
\end{widetext}

Using the distribution functions of Fig. \ref{fig:ln}(a), in Fig.
\ref{fig:ln}(b) we present the calculated averaged third-harmonic
response $\left\langle \left|\overline{M_{3}}\right|\right\rangle $
(solid red line) using the experimentally determined values for $T_{c}$
and $H_{c2}$. The comparison with the data shows that even a relatively
mild width of the distribution of $t_{c}$ values, with $\sigma\apprle0.1$,
is capable of capturing the extended temperature window for which
the third-harmonic response is sizable. As anticipated, this behavior
is a consequence of the fact that regions with a local higher $T_{c}$
value, although occupying a small volume, provide a sizable contribution
to the third-harmonic response.

The temperature dependence of the third-harmonic response data, however,
is not very well captured by the theoretical curves in Fig. \ref{fig:ln}(b).
To try to address this issue, we promote $T_{c}$ to a free parameter
and allow it to deviate slightly from the experimental value $T_{c}^{(\mathrm{exp})}=1.51$
K. Fig. \ref{fig:final} shows the results for $\left\langle \left|\overline{M_{3}}\right|\right\rangle $
in the case of $T_{c}=1.41\:\mathrm{K}\approx0.93T_{c}^{(\mathrm{exp})}$
and $\sigma=0.08$. Clearly, the temperature dependence of the calculated
$\left\langle \left|\overline{M_{3}}\right|\right\rangle $ becomes
more similar to the experimentally measured one, but still fails to
capture it completely. Thus, our conclusion is that while $T_{c}$
inhomogeneity may explain the extended temperature range where the
third-harmonic response is sizable, it is unlikely to explain the
exponential tail of $\left|\overline{M_{3}}\right|$ observed experimentally
in Ref. \citep{pelc2019}.

\section{Concluding remarks \label{sec:Conclusions}}

In this work, we used the LD model to compute the third-harmonic magnetic
response $\left|\overline{M_{3}}\right|$ due to Gaussian superconducting
fluctuations. Due to its phenomenological nature, the LD model could
in principle be applicable to both conventional and unconventional
superconductors. Our detailed comparison with measurements of $\left|\overline{M_{3}}\right|$
found that the theoretical modeling provides a good description of
the data in the case of Pb, Nb, and V -- provided that the critical
field is properly modified from its experimental value -- but a rather
poor account of the data for SRO. Inclusion of $T_{c}$ inhomogeneity,
which is intrinsically present in SRO, improved significantly the
agreement between theoretical model and experimental data, although
the model could not properly capture the experimentally observed exponential
temperature dependence of $\left|\overline{M_{3}}\right|$ (see Ref.
\citep{pelc2019}).

Further investigation is thus required to elucidate the origin of
this exponential behavior of $\left|\overline{M_{3}}\right|$, which
was also seen in other perovskite superconductors such as STO and
the cuprates, and appears to be quite robust \citep{pelc2019}. One
cannot completely discard simple $T_{c}$ inhomogeneity as the source
of this effect, since here we only focused on a very specific and
particularly simple distribution function for $T_{c}$. While this
choice allowed us to argue on a more quantitative basis that $T_{c}$
inhomogeneity can explain why $\left|\overline{M_{3}}\right|$ remains
large over a wide temperature window in SRO, the actual $T_{c}$ distribution
is certainly more complicated and likely material-dependent. A phenomenological
$T_{c}$ distribution will likely require fine tuning to give an exponential
temperature dependence of the third-harmonic response. Nevertheless,
if rare regions are present, they might give rise to specific tails
in the distribution function that may be common to different materials;
these types of effects have been explored in more detail in Refs.
\citep{dodaro2018,pelc2021}. We also note that, in the particular
case of the cuprates, an exponential temperature-dependent behavior
associated with superconducting fluctuations was also observed in
other observables such as linear/nonlinear conductivity and specific
heat, and described in terms of a Gaussian $T_{c}$ distribution \citep{popcevic2018,Pelc2018}.
It would be interesting to investigate whether the exponential temperature
dependence observed in the third-harmonic response of SRO is also
manifested in these other observables in the case of SRO. In fact,
as shown in Ref. \citep{pelc2019}, prior specific heat data \citep{nishizaki1999}
are consistent with this possibility.

Besides inhomogeneity, another effect, more specific to SRO, is that
if it is indeed a two-component superconductor, as proposed by different
models \citep{Romer2019,Agterberg2020,Kivelson2020,Willa2020}, the
superconducting fluctuation spectrum will likely be more complicated
than that of the LD model. However, the fact that the same exponential
temperature-dependence of $\left|\overline{M_{3}}\right|$ is seen
in STO and cuprates, the latter being single-component superconductors,
renders this scenario less likely. Finally, a crucial approximation
of the LD model is that it solely focuses on Gaussian superconducting
fluctuations. This raises the interesting question of whether non-Gaussian
fluctuations may also play an important role in the fluctuation spectra
of perovskite superconductors.
\begin{acknowledgments}
We thank Z. Anderson and S. Griffitt for assistance in ac susceptibility
probe design and construction, and A. Mackenzie and C. Hicks for providing
the SRO samples. This work was supported by the U. S. Department of
Energy through the University of Minnesota Center for Quantum Materials,
under Award No. DE-SC-0016371. 
\end{acknowledgments}

\bibliographystyle{apsrev}
\bibliography{diamagref}

\medskip{}

\end{document}